\documentclass[%
  jap,
  amsmath,%
  amssymb,%
  amsfont,%
  draft
  preprint,%
  author-year,%
  author-numerical,%
  superscriptaddress,%
  unsortedaddress	
]{revtex4-1}

\usepackage{graphicx}
\usepackage{color}
\usepackage{bm}

\usepackage{amssymb}

\usepackage{multirow}

\begin{document}




\title{Interface trap density metrology from sub-threshold transport in highly scaled undoped Si n-FinFETs}


\author{Abhijeet Paul} 
\email{abhijeet.rama@gmail.com}
\affiliation{School of Electrical and Computer Engineering, Network for Computational Nanotechnology, Purdue University, West Lafayette, 47907, USA.}
\author{Giuseppe C. Tettamanzi}
\email{g.tettamanzi@unsw.edu.au}
\affiliation{Kavli Institute of Nanoscience, Delft University of Technology, Lorentzweg 1, 2628 CJ Delft, The Netherlands and Centre for Quantum Computation and Communication Technology, University of New South Wales, Sydney NSW 2052, Australia.}
\author{Sunhee Lee, Saumitra R Mehrotra}
\affiliation{School of Electrical and Computer Engineering, Network for Computational Nanotechnology, Purdue University, West Lafayette, 47907, USA.}
\author{Nadine Collaert, Serge Biesemans}
\affiliation{IMEC, 3001 Leuven, Belgium.}
\author{Sven Rogge}
\affiliation{Kavli Institute of Nanoscience, Delft University of Technology, Lorentzweg 1, 2628 CJ Delft, The Netherlands and Centre for Quantum Computation and Communication Technology, University of New South Wales, Sydney NSW 2052, Australia.}
\author{Gerhard Klimeck}
\affiliation{School of Electrical and Computer Engineering, Network for Computational Nanotechnology, Purdue University, West Lafayette, 47907, USA.}

\pagenumbering{arabic}

\date{\today}

\begin{abstract}
Channel conductance measurements can be used as a tool to study thermally activated electron transport in the sub-threshold region of state-of-art FinFETs. Together with theoretical Tight-Binding (TB) calculations, this technique can be used to understand the evolution of source-to-channel barrier height ($E_{b}$) and of active channel area ($S$) with gate bias ($V_{gs}$). The quantitative difference between experimental and theoretical values that we observe can be attributed to the interface traps present in these FinFETs. Therefore, based on the difference between measured and calculated values of (i) $S$ and (ii) $|\partial E_{b}/\partial V_{gs}|$ (channel to gate coupling), two new methods of interface trap density ($D_{it}$) metrology are outlined. These two methods are shown to be very consistent and reliable, thereby opening new ways of analyzing in situ state-of-the-art multi-gate FETs down to the few nm width limit. Furthermore, theoretical investigation of the spatial current density reveal volume inversion in thinner FinFETs near the threshold voltage.  

\end{abstract}







\maketitle

\section{Introduction}
\label{intro_sec}

\begin{figure}[htb]
\centering
\includegraphics[width=3in,height=3in]{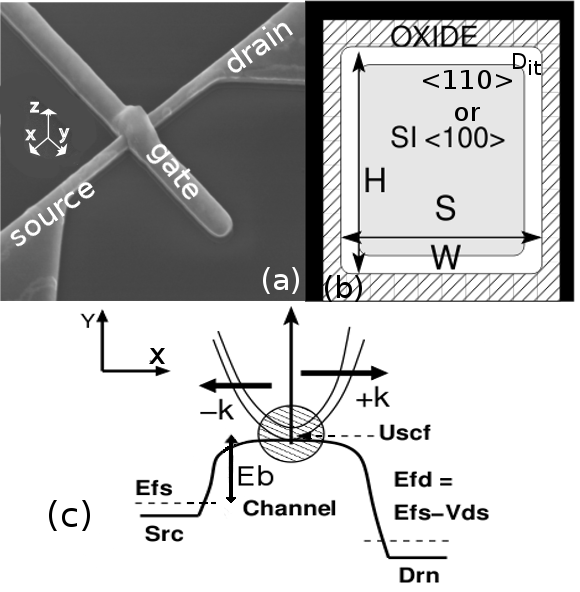}
\caption{(a) Scanning-electron-microscope (SEM) image of a Si n-FinFET with [100] channel orientation and single fin. (b) The schematic of the cross-sectional cut in
the Y-Z plane of a typical tri-gated FinFET. The active cross-section ($S$) is in gray, $H$ and $W$ are the physical height and width, respectively. (c) Ballistic top of the barrier model employed for calculating the thermionic current in the FinFETs.}
\label{fig:Finfet_cartoon}
\end{figure}

The non-planar tri-gated FinFET geometry (Fig.~\ref{fig:Finfet_cartoon}a) provides a viable solution to the channel length ($L_{ch}$) scaling due to its better gate to channel electrostatic coupling and reduced Short Channel Effects (SCEs) \cite{itrs,finfet1,finfet2,dev_detail,SCE_1}. In a recent experimental study of undoped Si n-FinFETs \cite{giuseppe}, thermionic emission in the sub-threshold region was used to measure (1) the active channel cross-section area (S) (Fig.~\ref{fig:Finfet_cartoon}b), which shows the region of channel where the charge prevalently flows and (2) the source to channel barrier height ($E_{b}$) (Fig.~\ref{fig:Finfet_cartoon}c), which reflects on the ease with which electrons travel from the source/drain to the channel. Understanding the evolution of the active area `S' and the barrier height `$E_{b}$' thus opens up new ways to investigate these FinFETs.

To shed light into the complicated transport phenomena that can arise in these undoped FinFETs we expand our previous work \cite{giuseppe,giuseppe_2} and  theoretically investigate the evolution of $S$ and $E_{b}$. Since these devices are small and have finite number of atoms in the channel, modeling transport requires an atomistic representation of the device. A 20 band $sp^3d^5s^*$ atomistic Tight-Binding (TB) model with spin orbit coupling (SO) \cite{TB_1,TB_2,Phytos_tob} is well suited for modeling the bandstructure of these confined Si channels, since TB can easily take into account the  material, geometrical, strain and potential fluctuations at the atomic scale \cite{Phytos_tob,OMEN}. This model also takes into account the coupling of the conduction (CB) and the valence bands (VB) which is neglected in simple models like the  effective mass approximation (EMA) \cite{jingwang}. Thermally activated transport is modeled using a semi-classical `Top of the barrier' (ToB) model (Fig. \ref{fig:Finfet_cartoon} c) \cite{Phytos_tob,paul_tob}. The simple ToB approach has been shown to model thermionic emission accurately \cite{kim_thermionic}.    

Qualitatively, we found similar theoretical and experimental trends for $S$ versus gate bias ($V_{gs}$) and $E_{b}$ versus $V_{gs}$ \cite{giuseppe}. However, the theoretically obtained $S$ and $E_{b}$ values quantitatively over-estimated the experimental values. The reduced experimental values can be attributed to the presence of interface traps in these FinFETs \cite{giuseppe,SCE_1,Kapila_trap,finfet_SRS_2}. The effect of interface traps on channel property are even more dominant in the extremely thin FinFETs \cite{giuseppe_2}. This difference in $S$ and $E_{b}$ has been utilized to directly estimate the interface trap density ($D_{it}$) in these FinFETs, thereby eliminating the need to implement special FinFETs geometries to determine $D_{it}$ \cite{Kapila_trap}. These methods now enable the direct implementation of interface trap density metrology in state-of-the-art undoped Si n-FinFETs. 

This paper has been divided into the following sections. Section \ref{sec_dev} provides the details about the FinFETs for which interface trap density metrology has been implemented and the fundamentals of our experimental procedures. The details about the self-consistent calculations are provided in Sec. \ref{sec_mod_self} and the theoretical extraction of $E_{b}$ and $S$ is outlined in Sec. \ref{sec_mod_Eb}. Section
\ref{extract_sec} provides the details of the two procedures for obtaining the interface trap density. The theoretical and experimental results and the discussion on them are given in Sec. \ref{sec_result}. The conclusions are summarized in Sec. \ref{sec_conc}.

\section{Device and Experimental details}\label{sec_dev}

\textit{Device details:} In this work 7 different FinFETs (labeled A-G) with two different channel orientations of $[100]$ ((FinFETs A-C and G)) and $[110]$ ((FinFETs D-F)) have been used \cite{dev_detail} (see Table \ref{dev_detail_table}). All the FinFETs have the same channel length ($L$ = 40nm). The channel height ($H$) is either 40nm or 65nm (Table \ref{dev_detail_table}). The channel width ($W$) varies between 3 to 25nm. All the FinFETs consist of one or more Si nanowire channels etched on a Si intrinsic film with a wrap-around gate covering three faces of the channels (Fig. \ref{fig:Finfet_cartoon} a) \cite{dev_detail}. An HfSiO (high-$\kappa$) layer isolates a TiN layer from the intrinsic Si channel \cite{dev_detail}. These FinFETs have either one channel (FinFETs A-C and G) or ten channels (FinFETs D-F). These devices have two different surface treatments (with or without $H_{2}$ annealing) as shown in Table \ref{dev_detail_table}. 

\textit{Measurement procedure:} The experimental value of $E_{b}$ and $S$ are obtained using a differential conductance ($G=\partial I_{D}/\partial V_{ds}$) method.
The conductance data are taken at $V_{ds}=0$ V using a lock-in technique. The full experimental method and the required ambient conditions have been outlined in detail in Ref.~\cite{giuseppe}.

\begin{table}[hbt]
\centering
\begin{tabular}{|c|l|l|l|c|l|}\hline
Label & H& W  & L & Channel & $H_{2}$  \\
      &[nm]   &[nm] & [nm]       & Orientation& anneal\\\hline
A  & 65 & 25 & 40 & [100] & Yes \\\hline
B & 65 & 25 & 40 & [100] & No \\\hline
C & 65 & $\sim$5 & 40 & [100] & No \\\hline
D  & 40 & 18 & 40 & [110] & Yes \\\hline
E & 40 & 18 & 40 & [110] & Yes \\\hline
F & 40 & $\sim$3-5 & 40 & [110] & Yes \\\hline
G & 65 & $\sim$7 & 40 & [100] & Yes \\\hline
\end{tabular}
\caption{Si n-FinFETs used in this study along with their labels. The surface hydrogen annealing detail is also shown. The channel is intrinsic Si, while the source and the drain are n-type doped for all the FinFETs.}
\label{dev_detail_table}
\end{table}

In the next section details of the theoretical approach to calculate the experimental values of $E_{b}$ and S in tri-gated n-FinFETs are outlined.

\section{Modeling Approach}
\label{modeling_sec}

\subsection{Self-consistent calculation}\label{sec_mod_self} 

The bandstructure for the Si channel is calculated using TB \cite{Phytos_tob,OMEN,paul_tob}. The TB calculation is coupled self-consistently to a 2D Poisson solver to obtain the charge and the potential \cite{Phytos_tob,paul_tob}. Once the convergence between the charge and the potential is achieved the thermionic current in the FinFETs is obtained using a semi-classical ballistic ToB model as shown in Fig.~\ref{fig:Finfet_cartoon}c \cite{Phytos_tob,paul_tob,virt_source}. Due to the extensively large cross-section of the devices that combines up to 44,192 atoms (for $H$ = 65nm, $W$ = 25nm FET) in the simulation domain, a new NEMO-3D code has been integrated in the top of the barrier analysis \cite{sunhee_tob}. Since the FinFETs studied here show (i) negligible source-to-drain tunneling current  and (ii) reduced SCEs \cite{giuseppe}, the ToB model is applicable to such devices \cite{paul_tob}. All the FinFETs are n-type doped in the source and drain to a value of 5$\times10^{19} cm^{-3}$. A 1.5nm SiO$_{2}$ cover is assumed.

Next we outline the procedure to calculate $E_{b}$ and $S$.

\subsection{Calculation of $E_{b}$ and S} \label{sec_mod_Eb}
For pure thermionic emission any carrier energetic enough to surmount the barrier from the source (Src) to the channel (C) (Fig.~\ref{fig:Finfet_cartoon}c) will reach the drain (Drn) provided the transport in the channel is close to ballistic \cite{virt_source}. Typically Src/Drn in FETs are close to thermal and electrical equilibrium (since heavy scattering in the contacts is assumed which leads to instantaneous carrier relaxation). This allows us to make the assumption that most of the carriers in the Src/Drn  are thermalized at their respective Fermi-levels ($E_{fs}$, $E_{fd}$ in Fig. \ref{fig:Finfet_cartoon} c). Also the channel potential ($U_{scf}$) can be determined under the application of $V_{gs}$ using the self-consistent scheme (discussed in Sec.\ref{sec_mod_self}). Hence, for the source-to-channel homo-junction inside a FET, the barrier height ($E_{b}$) can be determined as a function of $V_{gs}$,

\begin{equation}
\label{eqn_barht}
 E_{b}(V_{gs}) = U_{scf}(V_{gs}) - E_{fs}.
\end{equation}

This definition of $E_{b}$ implicitly contains the temperature dependence since the simulations are performed at different temperatures ($T$) which enters through the Fermi-Dirac distribution of the Src/Drc. We show in a later section, that the temperature dependence of $E_{b}$ in the sub-threshold region is very weak. Therefore, all the theoretical $E_{b}$ results shown in this work are at T = 300K.

The study of thermionic emission model is applicable when the barrier height is much larger than the thermal broadening ( $E_{b} \gg k_{B}T$ \cite{smsze}, where $k_{B}$ is the Boltzmann constant). For this reason, Eq. (\ref{eqn_barht}) works only in the subthreshold region where $E_{b}$ is well defined \cite{paul_tob}. Once the FinFET is above the threshold, $E_{b}$ ($\le k_{B}T$) is not a well defined quantity \cite{paul_tob}. Using the $E_{b}$ value, $S$ can be extracted using the conductance ($G$) in the thermionic emission regime for a 3D system \cite{smsze,giuseppe} as,  

\begin{equation}
\label{Gbulk}
   G_{3D} = SA^{*}T\frac{e}{k_{B}}exp\Big(-\frac{E_{b}(V_{gs})}{k_{B}T}\Big)
\end{equation}

where $A^*$ is the effective 3D Richardson constant ($A^{*}_{Si,3D} = 2.1 \times 120 \,A cm^{-2} K^{-2}$) \cite{smsze}, and $e$ is the electronic charge. This will hold only when the cross-section size of the FinFET is large enough (i.e.: $W$, $H$ $>$ 20nm) to be considered a 3D bulk system. For a very narrow width FinFET, $S$ cannot be extracted using (\ref{Gbulk}) since the system is close to 1D. For a 1D system the $G$, under a small drain bias ($V_{ds}$) at a temperature T, is given by the following relation (for a single energy  band \cite{G1D_derive}),

\begin{equation}
\label{G1D}
   G_{1D} = \frac{2e^2}{h}\cdot\Big[1+exp(\frac{E_{b}(V_{gs})}{k_{B}T})\Big]^{-1}
\end{equation}

where h is the Planck's constant. Since Eq.(\ref{G1D}) lacks any area description, G for 1D systems is no more a good method to extract $S$. 

\section{Trap extraction methods}
\label{extract_sec}

In Ref. \cite{giuseppe} it was observed that the active cross-section area ($S_{sim}$) obtained theoretically was an over-estimation of the experimental value ($S_{exp}$). In the results section it will be further shown that also the theoretical $E_{b}$ value can over estimate the experimental $E_{b}$ value. These mismatches can be attributed to the presence of traps at the oxide-channel interface of multi-gate FETs where these traps can enhance the Electrostatic-static screening and suppress the action of the gate on the channel \cite{Kapila_trap,finfet_SRS_2,giuseppe,giuseppe_2}. This simple idea is a powerful tool used for the estimation of interface trap density ($D_{it}$) in these undoped Si n-FinFETs. 

\subsection{Method I: $D_{it}$ from active area} \label{sec_ext_1}


Based on the difference between the simulated and the experimental active area ($S$) values, a method to calculate the density of interface trap charges, $\sigma_{it}$, in the FinFETs is outlined. The method is based on the assumption that the total charge in the channel at a given $V_{gs}$ must be the same in the experiments and in the simulations. This requirement leads to the following, 

\begin{equation}
\label{charge_equality}
S_{sim}\cdot L_{ch} \cdot \rho_{sim} = S_{expt}\cdot L_{ch} \cdot \rho_{expt} + e\cdot \sigma_{it}\cdot L_{ch} \cdot P
\end{equation} 

where $S_{sim}$ ($S_{expt}$) is the simulated (experimental) active area, $P$ is the perimeter of the channel under the gate ($P$ = $W$ + 2$H$) and $\rho_{sim}$ ($\rho_{expt}$) is the simulated (experimental) charge density. By applying Gauss's law at the oxide channel interface, $\rho_{expt}$ is obtained from $\rho_{sim}$ and $\sigma_{it}$ as, 

\begin{equation}
 \label{rhoexp} 
 \rho_{expt} = \rho_{sim} - \rho_{it} = \rho_{sim} - (e \cdot \sigma_{it} \cdot P)/(W\cdot H)
\end{equation}

Using (\ref{charge_equality}) and (\ref{rhoexp}) the final expression for $\sigma_{it}$ is obtained as,

\begin{eqnarray}
 \label{trap_den1}
\sigma_{it}(V_{gs}) & = & \frac{\rho_{sim}(V_{gs})S_{sim}(V_{gs})}{e \cdot P} \\ \nonumber
		   & &\times \left[\frac{\left[1-\frac{S_{expt}(V_{gs})}{S_{sim}(V_{gs})}\right]}{\left[1-\frac{S_{expt}(V_{gs})}{W\cdot H}\right]}\right] \;\; [\#/cm^2]
\end{eqnarray}

This method is useful for wider devices for which Eq.(\ref{Gbulk}) is valid. For very thin FinFETs (close to a 1D system) this method cannot be utilized .

\textit{Assumptions in Method I:} The extra charge contribution completely stems from the interface trap density ($D_{it}$) and any contribution from the bulk trap states has been neglected. All the interface traps are assumed to be completely filled which implies $\sigma_{it}$ $\cong$ $D_{it}$. The interface trap charges are assumed to be situated very close to the oxide-channel interface for Eq.\ref{rhoexp} to be true.  Also the interface trap density is assumed to be constant and identical for the top and the side walls of the FinFET which is generally not the case \cite{Kapila_trap,finfet_SRS_2}. This method of extraction works best for undoped channel since any filling of the impurity/dopant states is neglected in the calculation. Orientation dependent $D_{it}$ for different surfaces could be included as a further refinement.

\subsection{Method II: $D_{it}$ from barrier control }\label{sec_ext_2}

\begin{figure}[htb!]
\centering
\includegraphics[width=3in,height=2.2in]{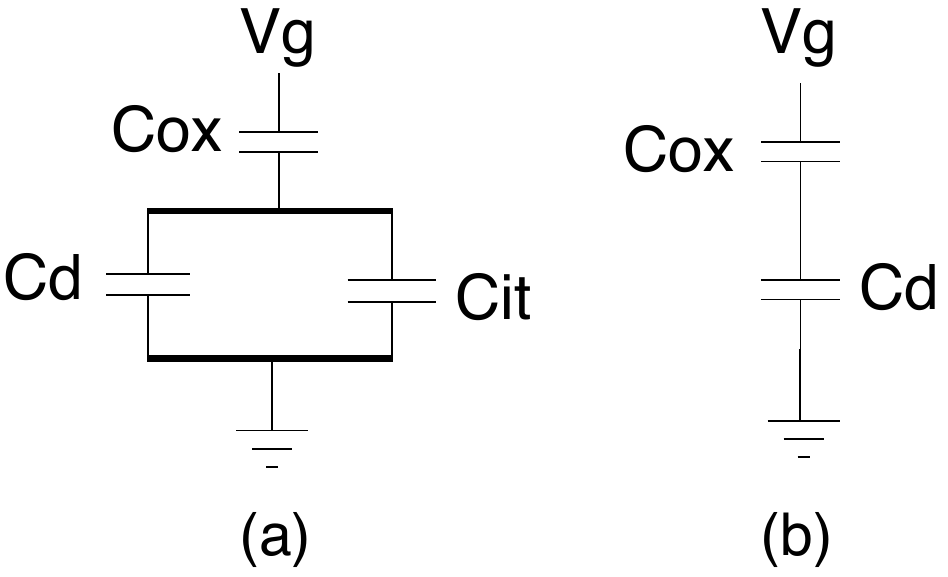}
\caption{Equivalent circuits (a) with interface-trap capacitance ($C_{it}$) and (b) without interface capacitance. $C_{d}$ and $C_{ox}$ are the depletion and the oxide capacitance, respectively. The idea for this equivalent circuit is obtained from page 381 in Ref. \cite{smsze}. }
\label{C_trap_fig}
\end{figure}

The second method does not utilize the $E_{b}$ value directly but its derivative w.r.t $V_{gs}$. The term $\alpha$ = $|\partial E_{b}/\partial V_{gs}|$ represents the channel to gate coupling \cite{giuseppe}. The presence of interface traps weakens this coupling due to the electrostatic screening. This method of trap extraction is based on the difference in the experimental and the simulated $\alpha$ value. The $\alpha$ value can be represented in terms of the channel and the oxide capacitance. The equivalent capacitance model for a MOSFET with and without interface traps ($D_{it}$) is shown in Fig.\ref{C_trap_fig}.

The $\alpha$ value can be associated to the oxide, interface and semiconductor capacitance which is given in Eq.(38) on page 383 in Ref. \cite{smsze}. This leads to the following relation,

\begin{equation}
\label{dEb_dEvg}
\vert \frac{\partial E_{b}}{\partial V_{gs}} \vert = 1 - \frac{C_{tot}}{C_{ox}},
\end{equation} 
where $C_{tot}$ is the total capacitance. For the two cases, as shown in Fig.\ref{C_trap_fig}, the total capacitance is given by, 
\begin{eqnarray}
\label{exp_c} C^{exp}_{tot} & = & \frac{C_{ox}\cdot (C_{d}+C_{it})}{C_{d}+C_{ox}+C_{it}}, \\
\label{sim_c} C^{sim}_{tot} & = & \frac{C_{d}\cdot C_{ox}}{C_{d}+C_{ox}},
\end{eqnarray} 

where $C_{it}$, $C_{ox}$ and $C_{d}$ are the interface trap capacitance, the oxide capacitance and the semi-conductor capacitance, respectively. Eq. (\ref{exp_c}) represents the capacitance in the experimental device and Eq. (\ref{sim_c}) represents the capacitance in the simulated device under ideal conditions without any interface traps. Combining Eq. (\ref{dEb_dEvg}), (\ref{exp_c}) and (\ref{sim_c}) and after some mathematical manipulations, we obtain,

\begin{equation}
\label{alpha_relation}
\frac{1}{\alpha_{exp}}  = \frac{1}{\alpha_{sim}} + \frac{C_{it}}{C_{ox}}, 
\end{equation}

Manipulating Eq. (\ref{alpha_relation}) gives the following relation for $C_{it}$,
\begin{equation}
\label{cit}
C_{it} =  C_{ox} \cdot \Big(\frac{1}{\alpha_{sim}}\Big)\cdot\Big[\frac{\alpha_{sim}}{\alpha_{exp}}-1\Big]
\end{equation}
Also $C_{it}$ can be related to the interface charge density ($\sigma_{it}$) as \cite{smsze},
\begin{equation}
\label{cit_2}
C_{it}  =  e \cdot \frac{\partial \sigma_{it}}{\partial V_{gs}}. 
\end{equation}

In Eq. (\ref{cit}) all the values are dependent on $V_{gs}$ except $C_{ox}$. Combining Eq. \ref{cit} and \ref{cit_2} and integrating w.r.t $V_{gs}$ yields the final expression for the integrated interface charge density in these FinFETs as,
\begin{eqnarray}
\label{method2_eq}
\sigma_{it}& = & \frac{C_{ox}}{e}\cdot \int_{V1}^{V2=V_{T}} \Big(\frac{1}{\alpha_{sim}(V_{gs})}\Big) \\ \nonumber
	   &   & \times \Big[\frac{\alpha_{sim}(V_{gs})}{\alpha_{exp}(V_{gs})}-1\Big] dV_{gs}\;\; [\#/cm^{2}],
\end{eqnarray}
where $V_{T}$ is the threshold voltage of the FinFET and V1 is the $V_{gs}$ at which $\alpha_{exp/sim} \approx 1$. Thus V1 and V2 is the integration range for Eq. (\ref{method2_eq}) in the sub-threshold region.

The second method derived from barrier control has the advantage that it is independent of the dimensionality of the FinFET. Hence, Eq. (\ref{method2_eq}) can be used for wide as well as thin FinFETs.

\textit{Assumptions in Method II:}  The  most important assumption is that the rate of change of the surface potential ($\Psi(V_{gs})$) is the same as $E_{b}$ w.r.t $V_{gs}$. The extra charge contribution completely originates from the density of interface trap charges ($\sigma_{it}$) and any contribution from the bulk trap states have been neglected. Also all the interface traps are assumed to be completely filled which implies $\sigma_{it}$ = $D_{it}$. This method works best when the change in the DC and the AC signal is low enough, such that the interface traps can follow the change in the bias sweep \cite{smsze} .

\subsection{Limitations of the methods}
It is important to understand the limitations of the new trap metrology methods to apply them properly. One of the main limitation is how closely the simulated FinFET structure resembles the experimental device structure. This depends both on the SEM/TEM imaging as well the sophistication of the model. In the present case we create the FinFET cross-section structure using the TEM image making the simulated structure as close to the experimental device as possible. With the development of better TCAD tools, the proximity of the simulated structure to experimental structure has increased. The physical device model needs to comprehend the crystal directions, atomistic details, strain and gating realistically to realize the working of the nano-scale FETs. Effective mass models fail to properly represent the bandstructure in these type of ultra-scaled nanowires/FinFETs \cite{jingwang}. Our model enables the calculation of theoretical conductance value with good confidence to be used in the trap calculation. Furthermore, the calculated G is calculated as close to ideal as possible and all the difference between the ideal and experimental G is attributed to the traps which may not be true always. An important difference between the two methods is that they are calculated over different $V_{gs}$ ranges. This is important since the trap filling and their behavior changes with $V_{gs}$ range which should be taken into account accurately.  One must also be aware of the embedded assumption of complete interface trap filling and the neglect of the bulk traps in the gate dielectric. 



\section{Results and Discussion}\label{sec_result}

In this section the theoretical results as well as their comparison with the experimental data are provided.

\subsection{3D vs. 1D system} \label{sec_3D_1D}

\begin{figure}[htb!]
\centering
\includegraphics[width=3in,height=2.4in]{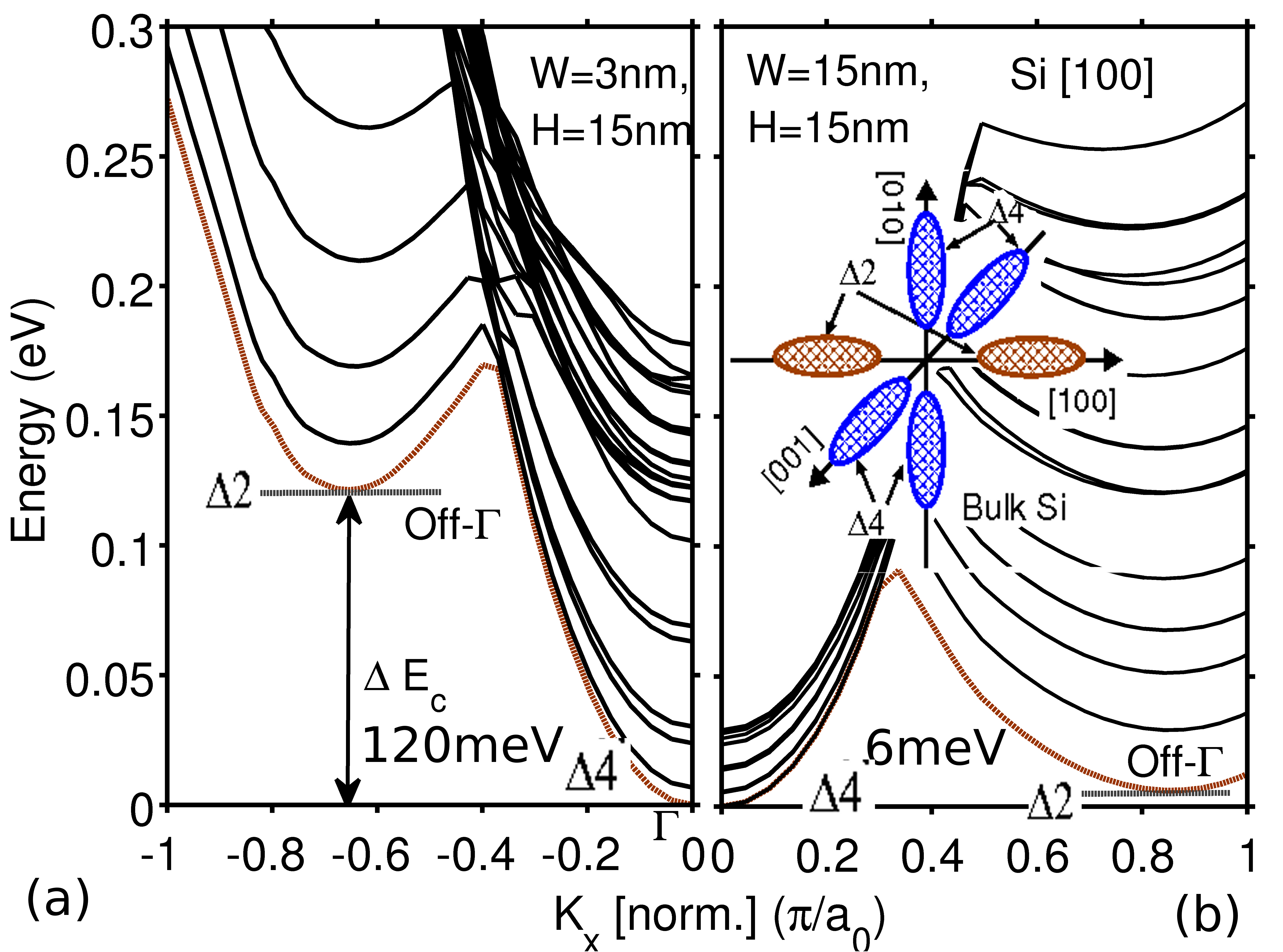}
\caption{Simulated conduction band E-K, using TB, for [100] Si channel with $H$ = 15nm for (a) $W$ = 3nm and (b) $W$ = 15nm. The inset shows 6 equivalent bulk Si conduction band ellipsoids. The $\Delta_{2}$ valleys (brown cigars) are along the transport direction [100] whereas the $\Delta_{4}$ valleys (blue cigars) are in the quantized plane.}
\label{Ek_sinw_fig}
\end{figure}

\begin{figure}[htb!]
\centering
\includegraphics[width=3in,height=2.4in]{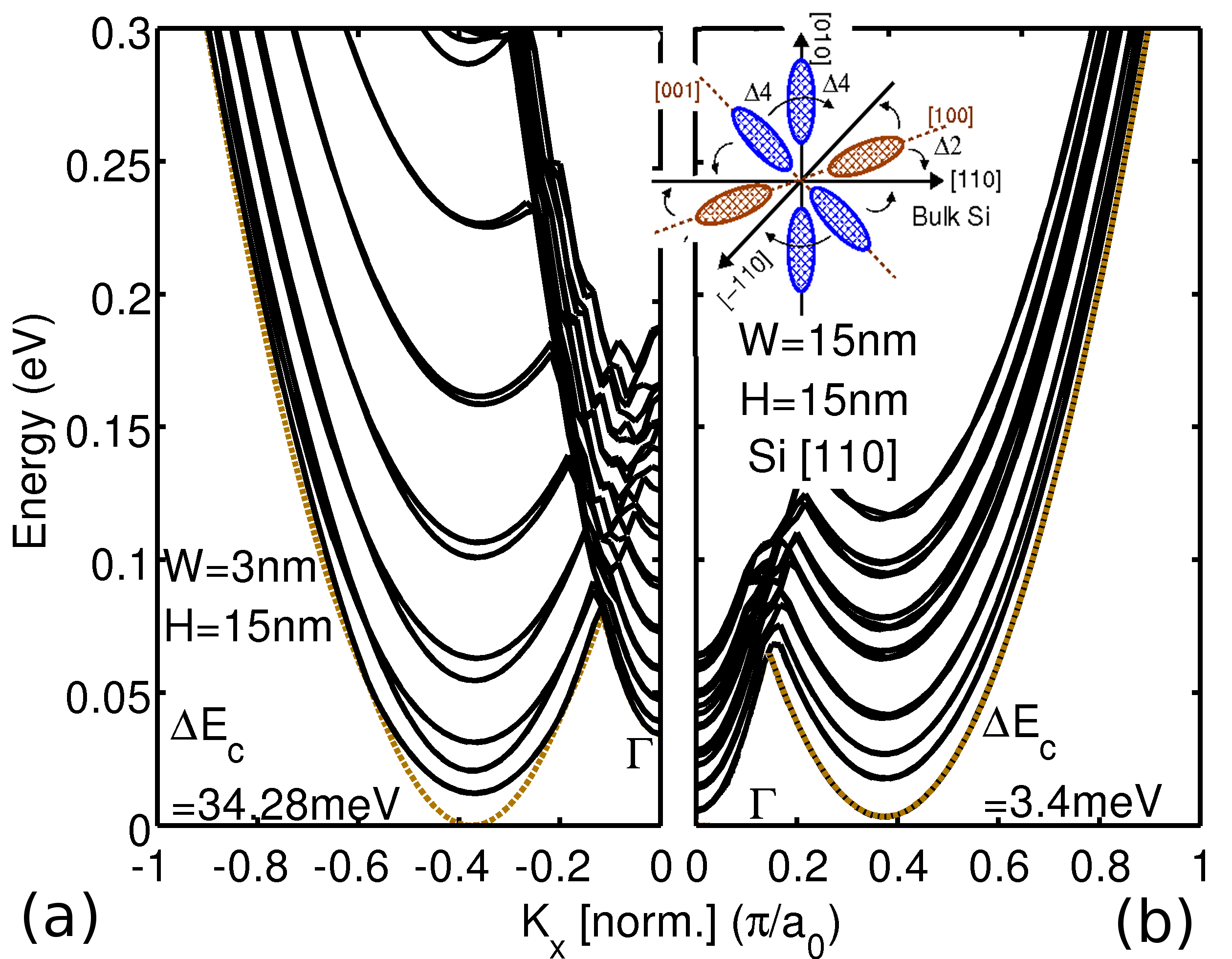}
\caption{Simulated conduction band E-K, using TB, for [110] Si channel with $H$ = 15nm for (a) $W$ = 3nm and (b) $W$ = 15nm. The CB minima is at Off-$\Gamma$ position for the thinner [110] Si channel. Inset shows the bulk Si 6 equivalent conduction valleys.}
\label{Ek_sinw_110_fig}
\end{figure}

The conduction band structure (E-K) can be utilized to distinguish a 1D system from a 3D system. The bulk silicon conduction band (CB) has 6 degenerate valleys ($\Delta_{6}$) (see inset of Fig. \ref{Ek_sinw_fig} and Fig. \ref{Ek_sinw_110_fig}) which split into 2 sets of degenerate valleys called the `$4-2\;configuration$' ($\Delta_{4}-\Delta_{2}$) for [100] and [110] 1D nanowires (NWs)  due to the geometrical confinement \cite{Phytos_tob}. In Si bulk the $\Delta_{2}$ valleys are along the [100] direction.  For a [100] Si nanowire channel the bulk $\Delta_{2}$ valleys are projected along the channel axis away from the $\Gamma$ point due to the folding of the Brillouin Zone (Fig. \ref{Ek_sinw_fig} a and b) \cite{Phytos_tob}. The bulk $\Delta_{4}$ valleys are projected at the $\Gamma$ point \cite{Phytos_tob}. The bandstructure of conduction band for silicon channel with $W$ = 3nm and $H$ = 15nm, and $W$ = $H$ = 15nm is shown in Fig. \ref{Ek_sinw_fig} a and b, respectively. For [110] oriented Si channel the valley projection is different compared to the [100] channel. The CB minima is at the Off-$\Gamma$ position as shown in Fig. \ref{Ek_sinw_110_fig}. This happens due to the different atomic positions and geometrical confinement in [100] and [110] channels \cite{Phytos_tob}.

The energy separation between the $\Gamma$ and Off-$\Gamma$ valleys is given by, 
\begin{equation}
\label{delEc_eq}
 \Delta E_{c} = E(\Gamma) - E(Off-\Gamma),
\end{equation} 
which gives a measure of how close (or far) a 1D NW system is from a 3D bulk system. The observation of a large $\Delta E_{c}$ value strongly points towards a 1D system, whereas a value close to zero points to a bulk system. 
Tight-binding simulations predict a $\Delta E_{c}$ value of around 120 meV for a [100] Si nanowire channel with $W$ = 3nm and $H$ = 15nm (Fig.\ref{Ek_sinw_fig} a) while this value reduces to 6 meV for a Si nanowires channel with $W$ = $H$ = 15nm (Fig. \ref{Ek_sinw_fig} b).  For a [110] Si channel the $\Delta E_{c}$ value is around 34 meV for $W$ = 3nm and $H$ = 15nm (Fig.\ref{Ek_sinw_110_fig} a) which reduces to 3.4 meV for $W$ = 15nm and $H$ = 15nm (Fig.\ref{Ek_sinw_110_fig} b). This indicates that larger cross-section silicon channels are closer to the 3D bulk system for both [100] and [110] oriented channels.

\begin{figure}[htb!]
\centering
\includegraphics[width=3.3in,height=1.8in]{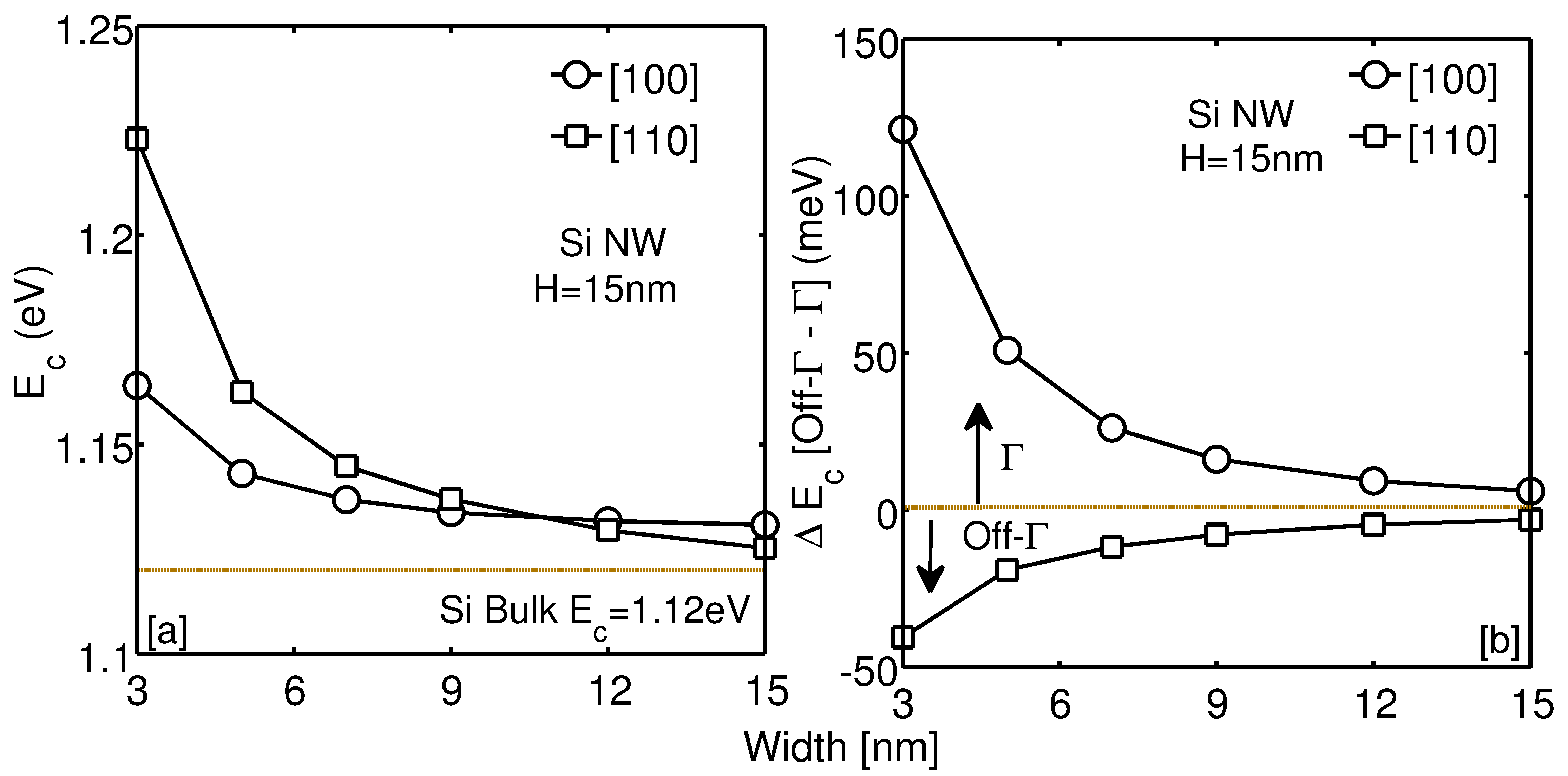}
\caption{ (a) Variation of the conduction band minimum ($E_{c}$) for [100] and [110] oriented Si channels with width for a fixed height of 15nm. (b) The separation of the $\Delta_{2}$-$\Delta_{4}$ valleys with width (W) for rectangular [100] and [110] Si channel for a fixed height of 15nm.}  
\label{EC_delEc_fig}
\end{figure}

The conduction band minimum (CBM) decreases with increasing channel width for both [100] and [110] SiNWs (Fig. Fig. \ref{EC_delEc_fig} a). Also  the $\Delta E_{c}$ value decreases with silicon channel width for a fixed height of 15nm ((Fig. \ref{EC_delEc_fig} b). The $\Delta E_{c}$ value is negative for [110] channel since the Off-$\Gamma$ valley is lower in energy compared to $\Gamma$ valley. For $W$ $>$ 15nm the $\Delta E_{c}$ is less than 5 meV ($\le k_{b}T_{300K}$) for both [100] and [110] Si channels.  \textit{This suggests that silicon channels with $W$ $\ge$ 15nm, and $H$ = 15nm behave electrically close to the bulk Si system at room temperature.} 

\subsection{Temperature dependence of $E_{b}$}
\begin{figure}[hbt!]
	\centering
		\includegraphics[width=3.0in,height=2.3in]{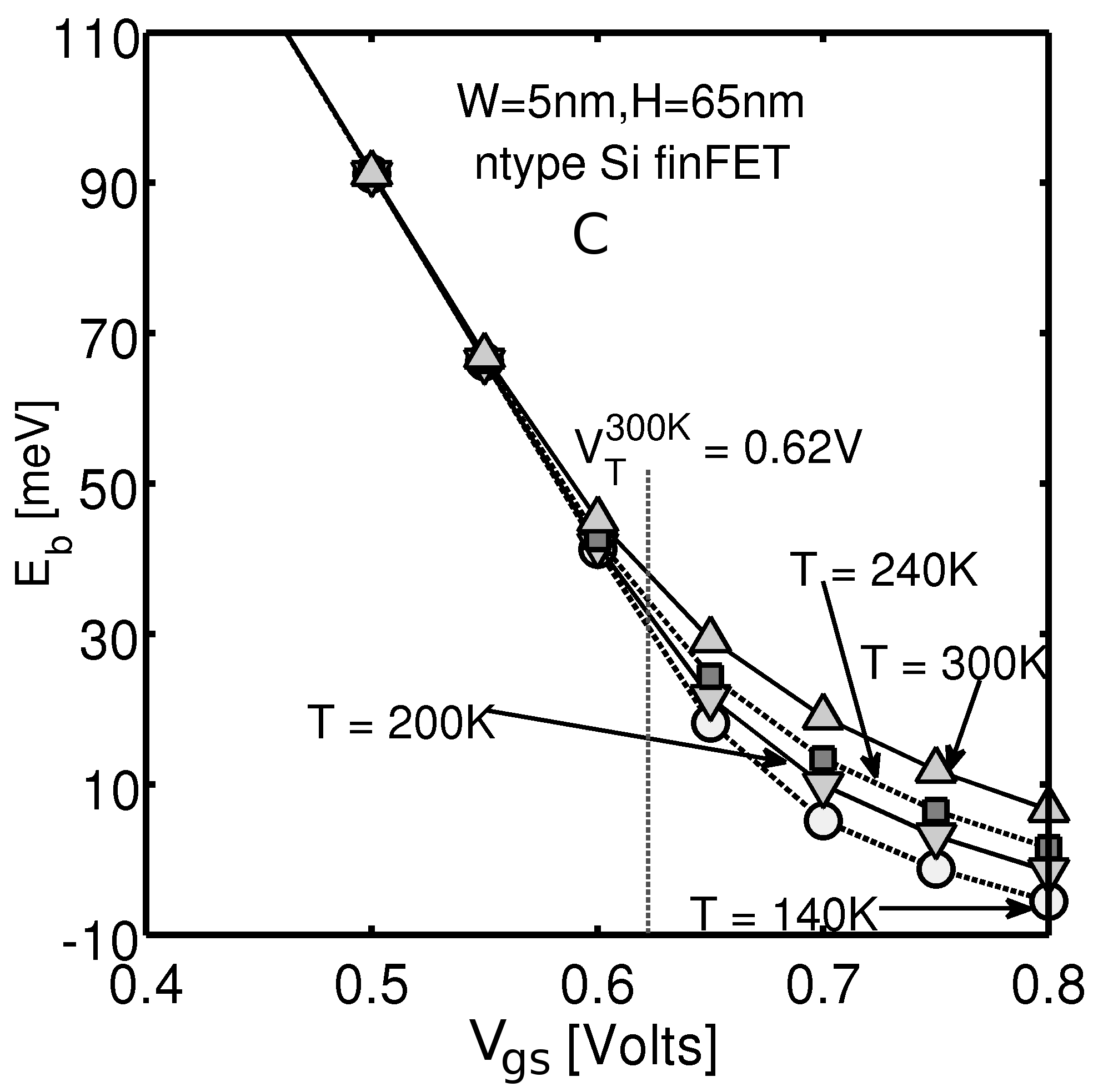}
	\caption{Temperature dependence of the simulated barrier height ($E_{b}$) in n-FinFET C. At T=300K, $V_{T}$ of the FinFET is 0.62V. The overlap of the curves at different temperatures with $V_{gs}$, below $V_{T}$ at 300K, shows a weak temperature dependence of $E_{b}$ in the sub-threshold region. The impact of temperature becomes prominent after $V_{gs}$ goes above $V_{T}$.}
	\label{fig:Eb_dev_E_temp}
\end{figure}

The source-to-channel barrier height has been assumed to be temperature independent in the sub-threshold region. Figure \ref{fig:Eb_dev_E_temp} shows the results of a temperature dependent ToB calculations and proves that the barrier height ($E_{b}$) is only weakly temperature dependent.
In the subthreshold region, the $E_{b}$ value for a device identical to FinFET C, is same at four different temperatures ( T=140K, 200K, 240K and 300K). The variation with temperature becomes more prominent when the FinFET transitions into the on-state. Thus, $E_{b}$ has a weak temperature dependence in the sub-threshold region allowing us to evaluate $E_{b}$ from the 300K simulations only. 

\subsection{Evolution of $E_{b}$ and S with $V_{gs}$} \label{sec_res_eb}

\begin{figure}[hbt!]
	\centering
		\includegraphics[width=3.4in,height=1.8in]{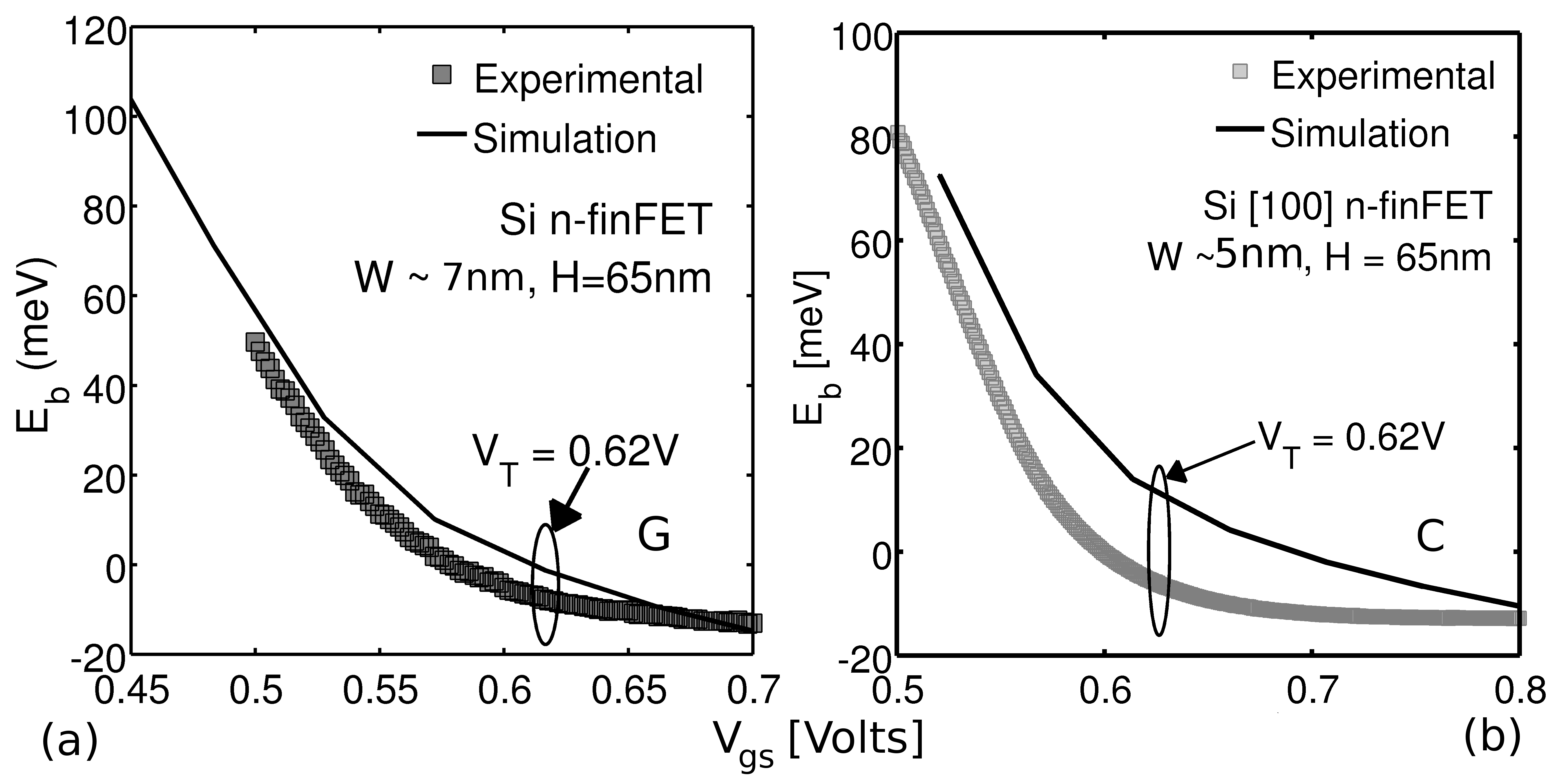}
	\caption{Experimental and simulated barrier height ($E_{b}$) in n-FinFET (a) G and (b) C. Both the devices have same $V_{T}$. Both experiment and 
	         simulation show a decreasing value of $E_{b}$ with $V_{gs}$, but the absolute values are different. }
	\label{fig:Eb_dev_C_D}
\end{figure}

\begin{figure}[hbt!]
	\centering
		\includegraphics[width=3.4in,height=1.8in]{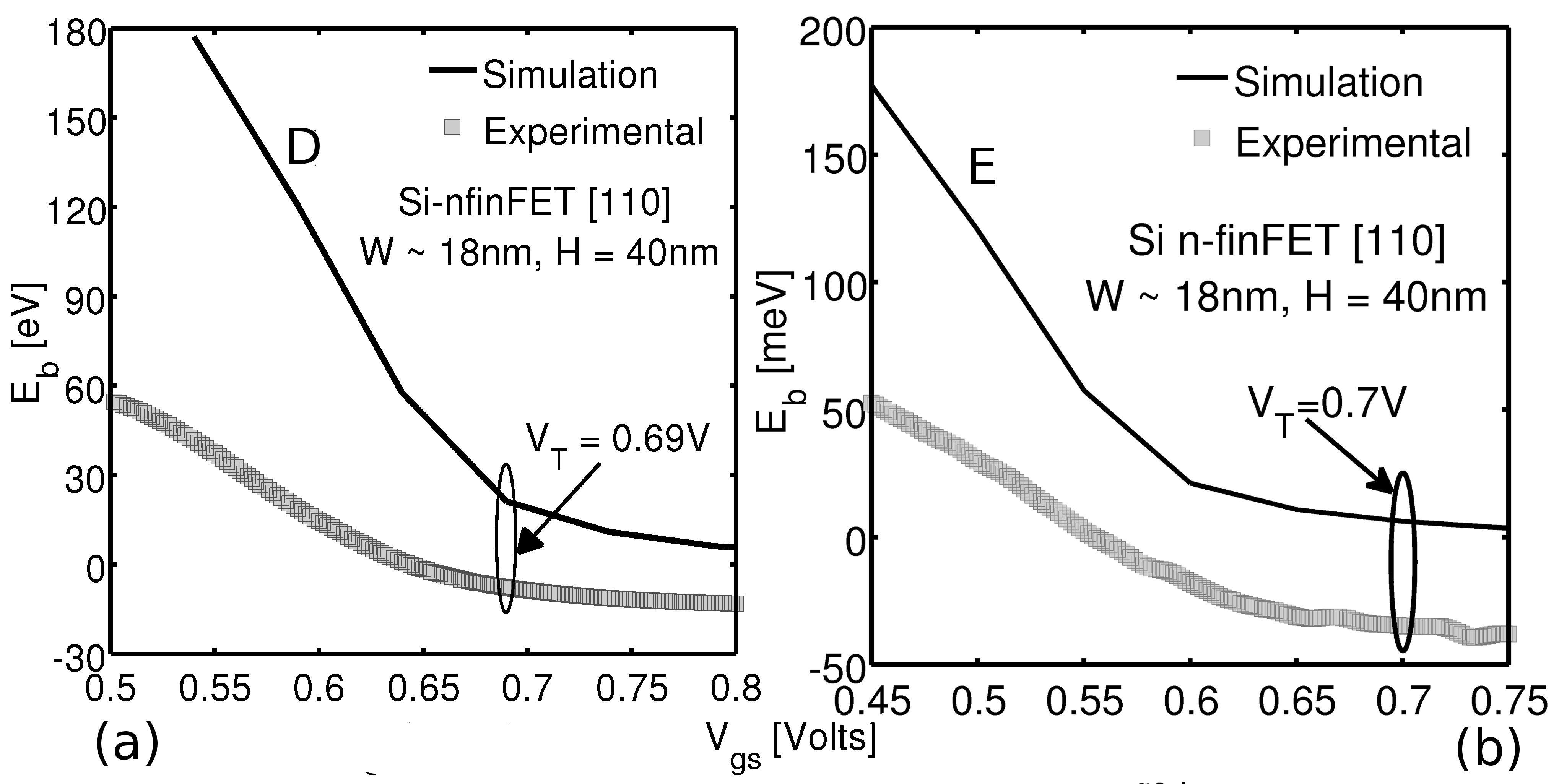}
	\caption{Experimental and simulated barrier height ($E_{b}$) in n-FinFETs (a) D and (b) E. Both the devices have similar $V_{T}$. Both experiment and 
	         simulation show a decreasing value of $E_{b}$ with $V_{gs}$, but the absolute values are different. }
	\label{fig:Eb_dev_E_F}
\end{figure}

Experimentally, it has been shown that, for undoped silicon n-FinFETs \cite{giuseppe}, $E_{b}$ reduces as $V_{gs}$ increases. Theoretically, the $E_{b}$ value is determined using Eq. (\ref{eqn_barht}) which depends on the self-consistent channel potential ($U_{scf}$). As the gate bias increases, the channel can support more charge. This is obtained by pushing the channel CB lower in energy to be populated more by the source and drain Fermi level \cite{Phytos_tob}. Figure~\ref{fig:Eb_dev_C_D} and \ref{fig:Eb_dev_E_F} show the experimental and theoretical evolution of $E_{b}$ in FinFETs G, C and D, E, respectively. Theory provides correct quanlitative trend for $E_{b}$ with $V_{gs}$. Few important observations here are, \textit{(i) the theoretical $E_{b}$ value is always higher than the experimental value and (ii) [110] Si devices (D and E) show a larger mismatch to the experimental values.} The reason for the first point is suggested to be the presence of interface traps in the FinFETs which screen the gate from the channel \cite{giuseppe,giuseppe_2}. The second observation can be understood by the fact that [110] channels with (110) sidewalls have higher interface trap density due to the higher surface bond density \cite{smsze} and poor anisotropic etching of the (110) sidewalls \cite{Kapila_trap,finfet_SRS_2}. 

\begin{figure}[hbt!]
	\centering
		\includegraphics[width=3.4in,height=1.8in]{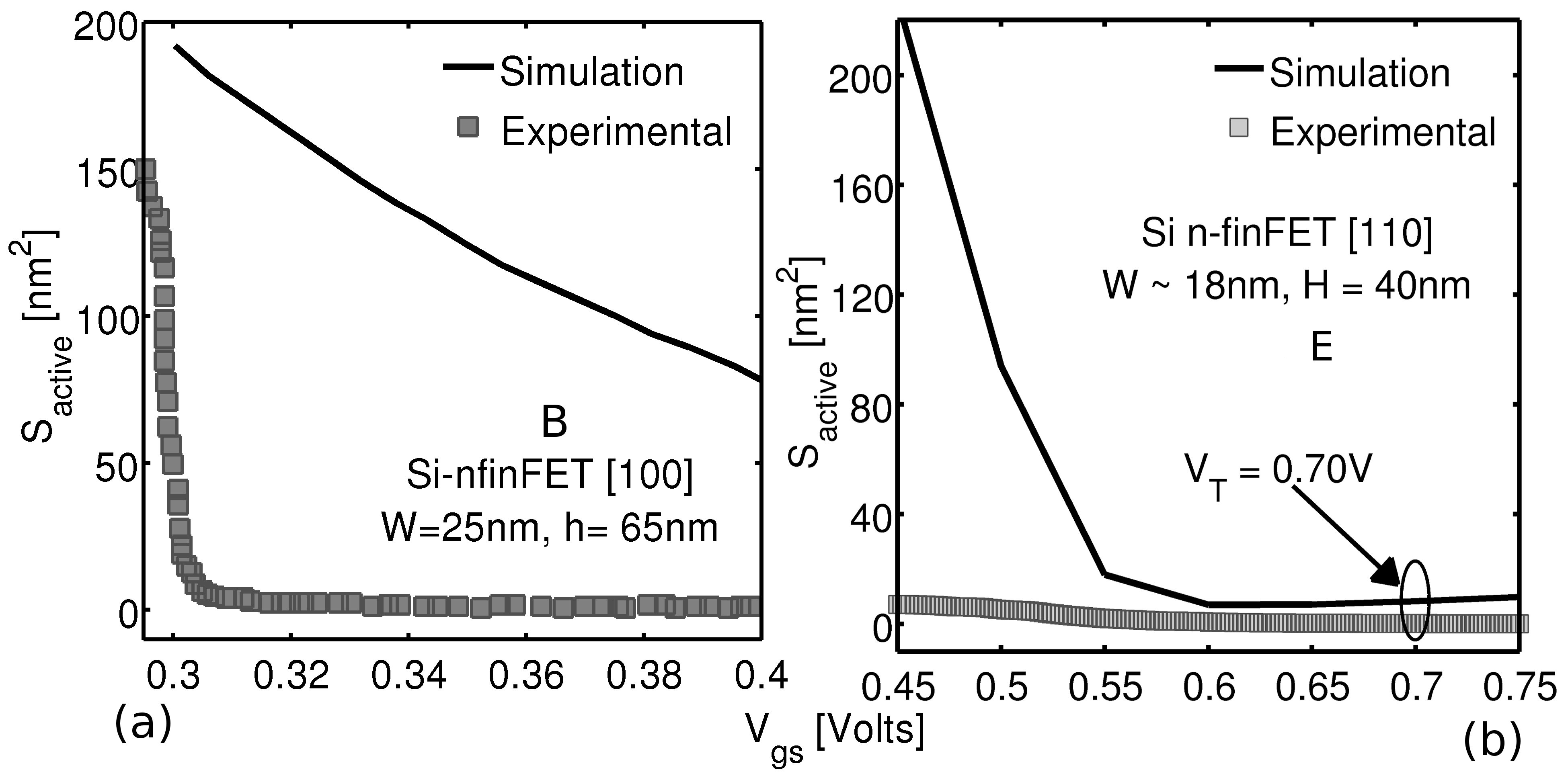}
	\caption{Experimental and simulated channel active cross-section (S) in n-FinFETs (a) B and (b) E. Both experiment and 
	         simulation show a decreasing value of $S$ with $V_{gs}$, but the absolute values are different. }
	\label{fig:S_dev_B_F}
\end{figure}

The active channel area ($S$) represents the part of the channel where the charge flows \cite{giuseppe}. Experimentally $S$ is shown to be decreasing with gate bias since the inversion charge moves closer to the interface which electrostatically screens the inner part of the channel from the gate \cite{giuseppe}. This gives a good indication of how much channel area is used for the charge transport. Figure \ref{fig:S_dev_B_F} a and b show the experimental evolution of $S$ in FinFET B and E, respectively. The theoretical value of $S$ decreases with $V_{gs}$ which is in qualitative agreement to the experimental observation \cite{giuseppe,giuseppe_2}. However, the absolute values do not match. In fact theory over-estimates the experimental $S$ value (Fig. \ref{fig:S_dev_B_F}) which is attributed to the interface traps. 

\subsection{Trap density evaluation}\label{sec_trap_res}
In this section we present the results on the interface trap density ($D_{it}$) in the undoped Si n-FinFETs.

\subsubsection{$D_{it}$ using $S$: Method I} 
\begin{figure}[hbt!]
	\centering
		\includegraphics[width=3.4in,height=1.8in]{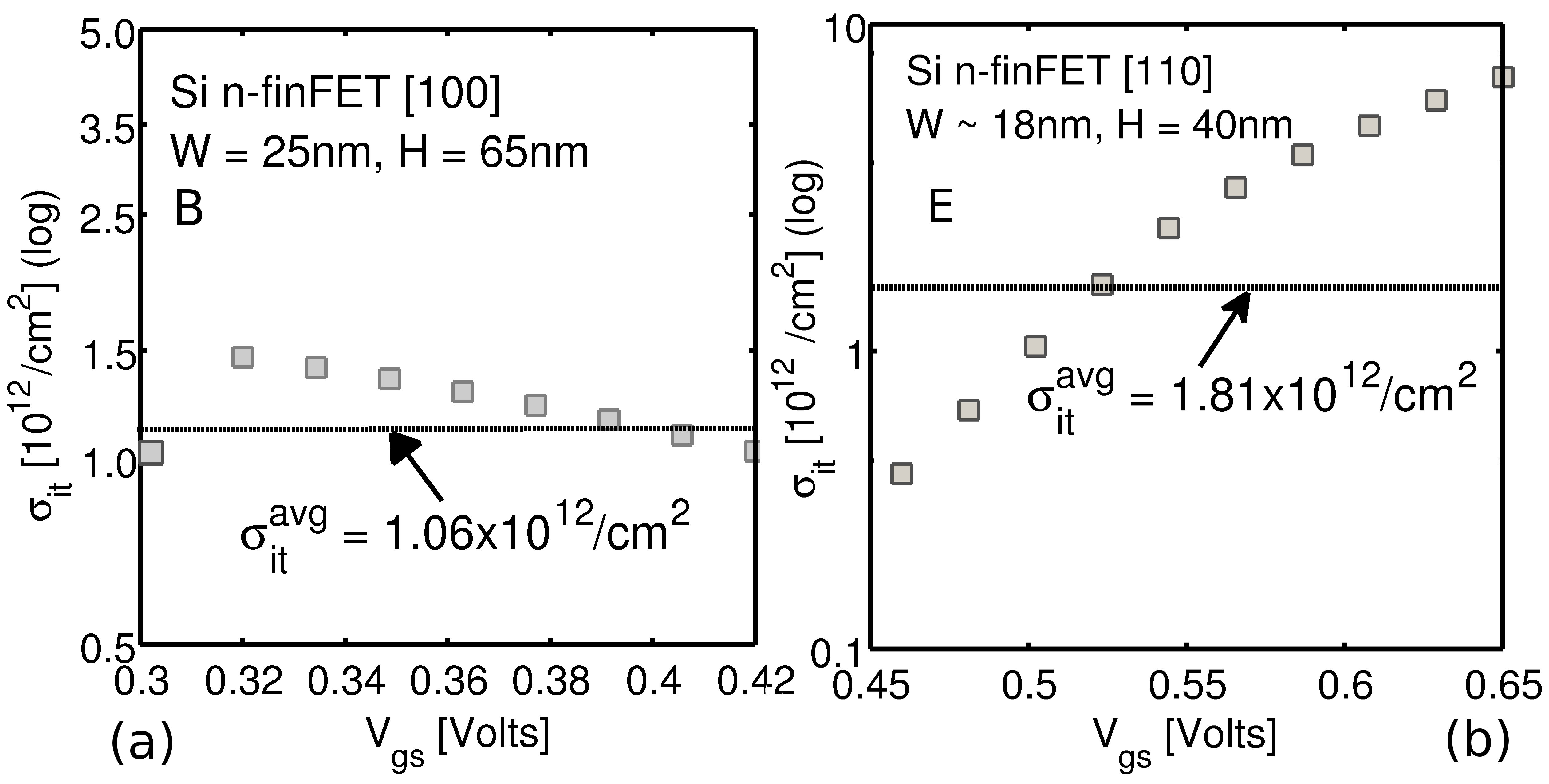}
	\caption{Extracted trap density using the difference in active device area (method I) for n-FinFETs (a) B and (b) E.}
	\label{fig:S_dit_B_F}
\end{figure}

The calculated $D_{it}$ values for FinFET B and E are 1.02e12$cm^{-2}$ and 1.81e12$cm^{-2}$ (Fig.\ref{fig:S_dit_B_F} a and b, respectively). The $D_{it}$ values compare quite well with the experimental $D_{it}$ values presented in Ref.\cite{Kapila_trap} and also shown in Table \ref{trap_den_table}. As expected the $D_{it}$ value for FinFET E (with [110] channel and (110) sidewalls) is higher than FinFET B ([100] channel with (100) sidewalls). This is attributed to the higher $D_{it}$ ($\sim 2\times$) on the (110) surfaces \cite{Kapila_trap,finfet_SRS_2}. Our results show $\sim 1.8\times$ more $D_{it}$ for (110) sidewalls, in close agreement to the experiments. This method allows to obtain the $D_{it}$ in the actual FinFETs rather than custom made FETs.

\subsubsection{$D_{it}$ using $|\partial E_{b}/\partial V_{gs}|$: Method II} 
\begin{figure}[hbt!]
	\centering
		\includegraphics[width=3.4in,height=1.8in]{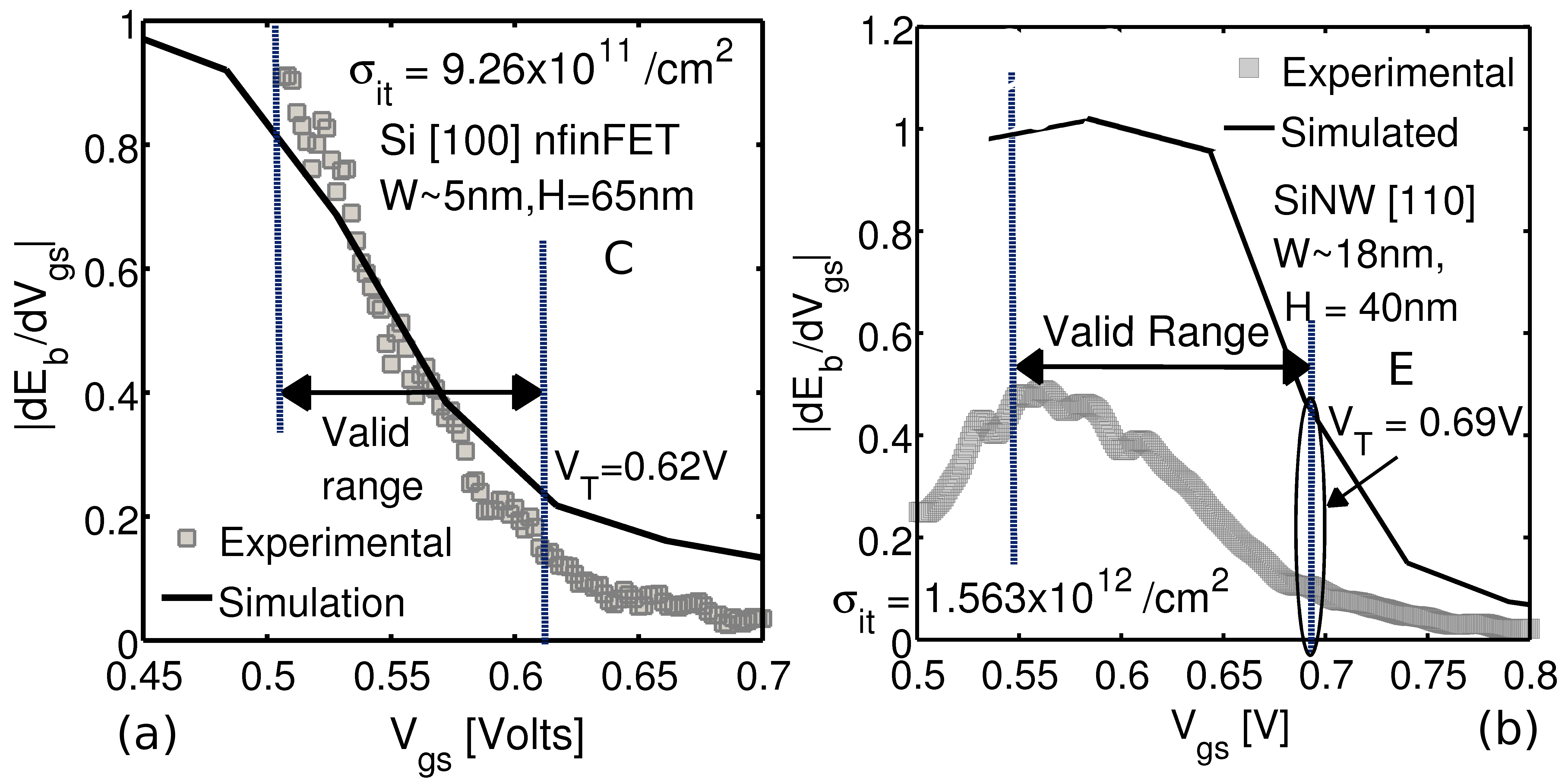}
	\caption{Experimental and simulated value of $\alpha$ in n-FinFETs (a) C and (b) E.}
	\label{fig:dEb_dvg_C_E}
\end{figure}

The $C_{ox}$ value needed in this method is taken as $\sim$0.0173 $F/m^{2}$ which is assumed to be the same for all the devices since these FinFETs have similar oxide thickness. The calculated $D_{it}$ values for FinFET C and D are 9.26e11$cm^{-2}$ and 1.563e12$cm^{-2}$ (Fig.\ref{fig:dEb_dvg_C_E} a and b, respectively). These calculations also show that the [110] channel device (FinFET E) shows a higher $D_{it}$ compared to the [100] channel device (FinFET C), again consistent to the observations made in Ref. \cite{Kapila_trap}. The advantage of this method is that it can be used to obtain $D_{it}$ in extremely thin FinFETs (close to 1D system) unlike method I which is applicable only to wider FinFETs (due to the reasons discussed in Sec. \ref{sec_mod_Eb}).  

\subsubsection{Discussion on the two methods and $D_{it}$ trends} 
The $D_{it}$ values for all the FinFETs used in this study are shown in Table \ref{trap_den_table}. The important outcomes about the two methods are outlined below:
\begin{itemize}
\item{The $D_{it}$ values obtained by the two methods compare very well with the experimental measurement in Ref.\cite{Kapila_trap} for similar sized FinFETs (A and B), demonstrating the validity of these new methods.}
\item{The $D_{it}$ values calculated using method I and II (for B and E) compare very well with each other which shows that the two methods are complimentary \cite{giuseppe_2} for large cross-section devices.}
\item{The $D_{it}$ values calculated for the two similar FinFETs (D and E) compare very well showing the reproducibility of the methods \cite{giuseppe_2}.}
\end{itemize}

The calculated $D_{it}$ values also reflect some important trends about the FinFET width scaling and surfaces (Table \ref{trap_den_table}). The central points are :
\begin{itemize}
\item{Hydrogen passivation considerably reduces $D_{it}$ \cite{SCE_1}. This is observed for FinFETs A and B where $H_{2}$ passivation results in $\sim 2 \times$ less $D_{it}$ in FinFET A \cite{giuseppe_2}}.
\item{Width scaling requires more etching which also increases $D_{it}$ \cite{Kapila_trap,finfet_SRS_2}. The same trend is observed in devices A to C and D to F (decreasing W).}
\item{(110) sidewalls show higher $D_{it}$ compared to (100) sidewalls \cite{Kapila_trap}. The same trend is also observed for FinFETs A, B, C, G ((100) sidewall) compared to FinFETs D, E and F ((110) sidewall).} 
\end{itemize}

\begin{table*}[hbt]
\centering
\begin{tabular}{|l|c|c|c|c|}\hline
Device & Method & $D_{it}\;(10^{11}cm^{-2})$& FET type & Observations \\\hline

L=140nm \cite{Kapila_trap} & Charge  & 1.725 &Special body  &  -- \\\cline{1-1} \cline{3-3} \cline{5-5}  
L = 240nm \cite{Kapila_trap} & Pumping & 2.072 & tied FET & -- \\\hline
A & I &  5.560 & Std. FET & $H_{2}$\\\cline{1-4}
\multirow{2}{*}{B} & I  & 10.60 & Std. FET & anneal,  \\\cline{2-4}
 	           & II & 8.860 & Std. FET & reduces $D_{it}$ \\\hline
C 		   & II & 9.26 &  Std. FET & Thin fin, more $D_{it}$  \\\hline

D                  & II & 18.31 & Std. FET& (110) side-wall, \\\cline{1-4}
\multirow{2}{*}{E} & I  & 18.1 & Std. FET &  thin fin, \\\cline{2-4}
 		   & II & 15.3 & Std. FET &  more etching, \\\cline{1-4}
F 		   & II & 36.3 & Std. FET &  more $D_{it}$ \\\hline
G 		   & II & 4.33 & Std. FET  & $H_{2}$ anneal, less $D_{it}$ \\\hline
\end{tabular}
\caption{Values of $D_{it}$ obtained from all the n-FinFETs as well as from Ref.\cite{Kapila_trap}.}
\label{trap_den_table}
\end{table*}

\subsection{Current distribution}\label{sec_current_res}

\begin{figure}[htb]
	\centering
		\includegraphics[width=3.1in,height=1.85in]{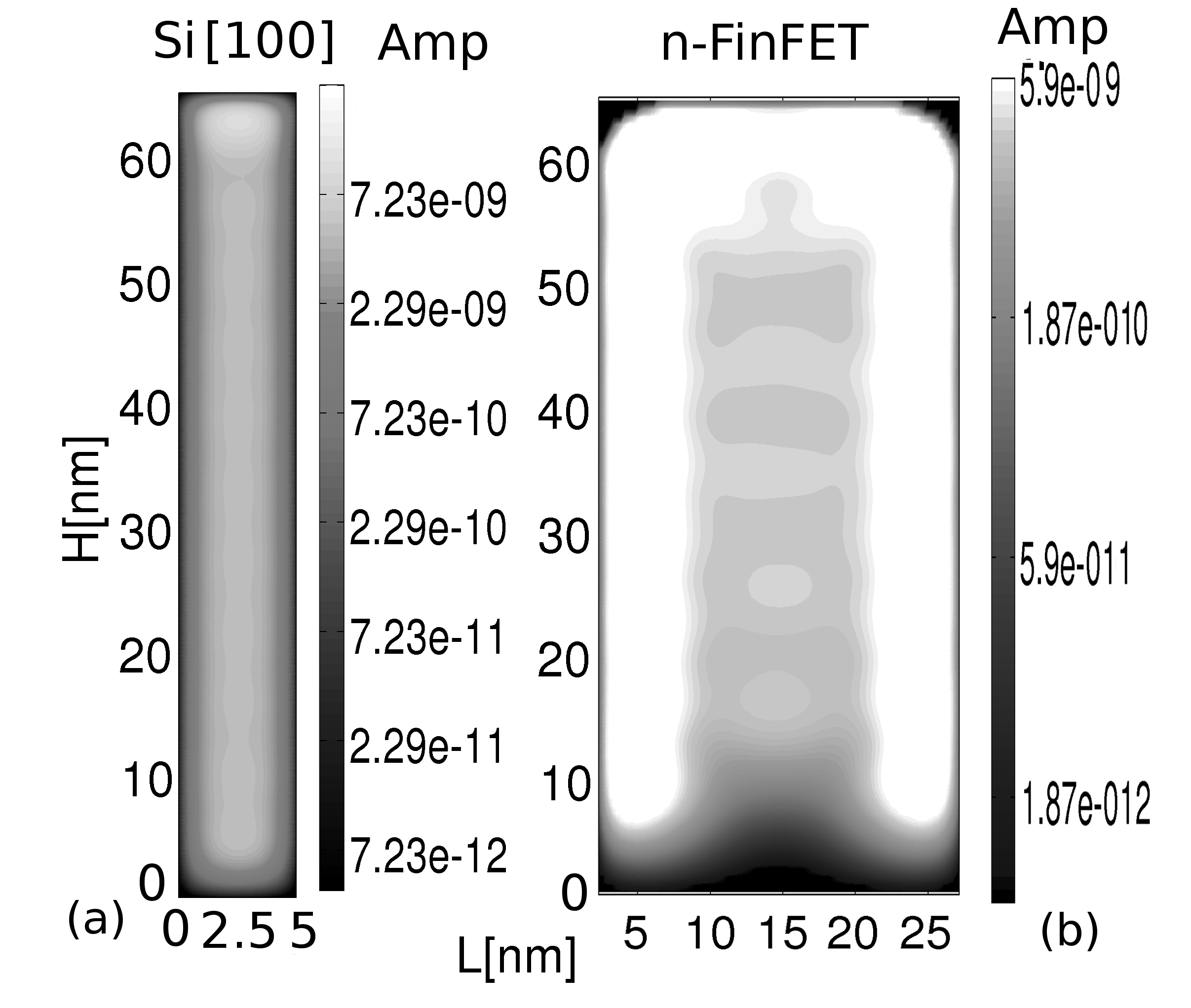}
	\caption{Spatial current distribution in the $\langle 100 \rangle$ undoped Si n-FinFET intrinsic with H = 65nm and (a) W = 5nm and (b) W = 25nm. $V_{gs}$ = 0.4V and $V_{DS}$ = 30mV at 300K. 5nm device shows a complete volume inversion. In the 25nm device the current mainly flows at the edges.}
	\label{fig:curr_dist}
\end{figure}

The spatial current distribution in FinFETs can provide a better insight for optimizing the channel cross-section area utilization. Theoretical calculations show that the charge flow in n-FinFETs depends strongly on the geometrical confinement. For very small width FinFET (W $\sim$ 5) the entire body gets inverted (Fig.\ref{fig:curr_dist} a) and shows little change in $S$ with $V_{gs}$ \cite{giuseppe}. For wider FinFETs (W $\sim$ 25) the current flow starts from a weak volume inversion and moves towards surface inversion as $V_{gs}$ increases (Fig.\ref{fig:curr_dist} b) \cite{giuseppe}. 
For extremely thin n-FinFETs ($W$ = 5nm, $H$ = 65nm) the charge flows through the entire body (volume inversion) compared to the wider n-FinFETs ($W$ = 25nm, $H$ = 65nm) where the charge flows at the edges. Thus thin FinFETs can show better channel area utilization for the charge flow compared to the wider devices. However thin FinFETs show an increase in $D_{it}$ due to more side-wall etching (Table \ref{trap_den_table}) which can severely limit the action of the thin FinFETs. The advancement of fabrication methods and strain technology may improve the performance of thin FinFETs as reported by some experimental works \cite{finfet_5,finfet_orient_1,finfet_2}. 

\section{Conclusions}\label{sec_conc}
Two new trap charge density metrology methods in ultra-scaled Si n-FinFETs are presented. The Top-of-the-barrier model, combined with Tight-binding calculations, explains very well the thermally activated sub-threshold transport in state-of-the-art Si FinFETs. The qualitative evolution of $E_{b}$ and S with $V_{gs}$ are well explained by the theory. The systematic mismatch in the experimental and theoretical values of $E_{b}$ and $S$ led to the development of two new interface trap density metrology schemes. The advantage of these schemes is that they do not require any special MOSFET structure as needed by the present experimental methods allowing to probe the interface quality of the ultimate channel. These methods are shown to provide consistent and reproducible results which compare very well with the independent experimental trap measurement results. The calculated trends of interface trap density with channel width scaling, channel orientation and hydrogen passivation of the surfaces show a good correlation with the experimental observations. Thin width FinFETs could lead to a better channel utilization due to strong volume inversion only if surface roughness and the density of interface traps created during the extreme etching of these ultra-scaled devices can be reduced.

\section*{Acknowledgment}
A.P, S.L, S.R.M and G.K would like to acknowledge the financial support from FCRP-MSD under Semiconductor Research Corporation, Nano Research Initiative (NRI) under Midwest Institute for Nanoelectronics Development (MIND) and National Science Foundation (NSF). Computational resources provided by nanoHUB.org, funded by NSF under Network for Computational Nanotechnology (NCN), is also acknowledged. G.C.T. and S.R. acknowledge FOM and the European Community Seventh Framework under the Grant Agreement nr: 214989-AFSiD for the financial support and the Australian Research Council Centre of Excellence for Quantum Computation and Communication Technology (project number CE110001029). G.C.T. acknowledge the kind hospitality extended by Prof. A. Di Carlo at the University of Tor Vergata, Rome, during the preparation of this manuscript and D. Brousse for the help in the acquisition of the SEM images. All the authors acknowledge B. Johnson, J. McCallum and N. Zimmerman for useful discussions. The band structure calculations for the silicon channels are done using Bandstructure Lab on nanoHUB.org \cite{bslab}. A.P. and G.C.T. contributed equally to this work. 


%



\bibliography{refs}







\end{document}